\documentclass{sigchi}

\usepackage{balance}  
\usepackage{graphics} 
\usepackage{txfonts}
\usepackage{times}    
\usepackage[pdftex]{hyperref}
\usepackage{color}
\usepackage{textcomp}
\usepackage{booktabs}
\usepackage{ccicons}
\usepackage{todonotes}

\usepackage{algorithm}
\usepackage{algorithmic}
\usepackage{multirow}
\usepackage{url}

\usepackage{array}
\newcommand{\PreserveBackslash}[1]{\let\temp=\\#1\let\\=\temp}
\newcolumntype{C}[1]{>{\PreserveBackslash\centering}p{#1}}
\newcolumntype{R}[1]{>{\PreserveBackslash\raggedleft}p{#1}}
\newcolumntype{L}[1]{>{\PreserveBackslash\raggedright}p{#1}}

\makeatletter
\def\url@leostyle{%
  \@ifundefined{selectfont}{\def\UrlFont{\sf}}{\def\UrlFont{\small\bf\ttfamily}}}
\makeatother
\def\pprw{8.5in}
\def\pprh{11in}

\definecolor{linkColor}{RGB}{6,125,233}
\hypersetup{%
  pdftitle={SIGCHI Conference Proceedings Format},
  pdfauthor={LaTeX},
  pdfkeywords={SIGCHI, proceedings, archival format},
  bookmarksnumbered,
  pdfstartview={FitH},
  colorlinks,
  citecolor=black,
  filecolor=black,
  linkcolor=black,
  urlcolor=linkColor,
  breaklinks=true,
}

\begin{document}

\title{Understanding Tree: a tool to estimate one's understanding of knowledge}

\numberofauthors{1}
\author{%
	\alignauthor{Gangli Liu\\
		\affaddr{Tsinghua University}\\
		\affaddr{Beijing, China}\\
		\email{gl-liu13@mails.tsinghua.edu.cn}}\\
}

\maketitle

\begin{abstract}
	People learn whenever and wherever possible, and whatever they like or encounter--Mathematics, Drama, Art, Languages, Physics, Philosophy, and so on. With the bursting of knowledge, evaluation of one's possession of knowledge becomes increasingly difficult. There are a lot of demands to evaluate one's understanding of a piece of knowledge. Assessment of understanding of knowledge is conventionally through tests or interviews, but they have some limitations such as low-efficiency and not-comprehensive. This paper proposes a method called Understanding Tree to estimate one's understanding of knowledge, by keeping track of his/her learning activities. It overcomes some limitations of traditional methods, hence complements traditional methods.
\end{abstract}

\keywords{Knowledge evalutaion; Understanding Tree; Familiarity Measure; Knowledge Model.}

\category{H.5.m.}{Information Interfaces and Presentation
  (e.g. HCI)}{Miscellaneous} \category{H.5.3.}{Group and Organization Interfaces}{Computer-supported cooperative work}

\section{1. Introduction}

Our world is bursting with knowledge. People's learning of knowledge is not confined to childhood or the classroom but takes place throughout life and in a range of situations; it can take the form of formal learning or informal learning\cite{Paradise2009}, such as daily interactions with others and with the world around us. People learn whenever and wherever possible. Lifelong learning is the "ongoing, voluntary, and self-motivated" pursuit of knowledge for either personal or professional reasons\cite{Cliath2000}. According to Tough's study, almost 70\% of learning projects are self-planned\cite{Tough1979}. 
\subsection{Definition of knowledge and learning}
Knowledge is conventionally defined as beliefs that are true and justified. To be `true' means that it is in accord with the way in which objects, people, processes and events exist and behave in the real world. However, exactly what evidence is necessary and sufficient to allow a true belief to be `justified' has been a topic of discussion (largely among philosophers) for more than 2000 years\cite{hunt2003concept}.

Learning is the process of acquiring, modifying, or reinforcing knowledge, behaviors, skills, values, or preferences in memory \cite{terry2015learning, mangal2002advanced, hunt2003concept, bransford1999people}. An individual's possessing of knowledge is the product of all the experiences from the beginning of his/her life to the moment at hand\cite{hunt2003concept,benassi2014applying}. Learning produces changes in the organism and the changes produced are relatively permanent\cite{schacter2011psychology}.
\subsection{Evaluation of one's possession of knowledge}
As people learn eternally, one critical question is how to evaluate how much knowledge has been possessed by an individual at some time. At present, assessment of one's possession of knowledge is primarily through tests\cite{pisa2000measuring,hunt2003concept,qian2004evaluation} or interviews.

In \cite{Liu2016}, some use cases of evaluating one's possession of knowledge are proposed. In addition, a method named Knowledge Model is devised to evaluate one's possession of knowledge, by keeping track of his learning activities. In Knowledge Model, knowledge is segmented as Knowledge Points and organized into a tree structure. A Knowledge Point is a piece of knowledge which is explicitly defined and has been widely accepted.

A system continually records the starting and cessation time of each learning activity, separating the learning activities into a series of learning sessions. Meanwhile, the text content of each learning session is extracted, then topic model is used to analyze the ingredients of the text content. After topic model analysis, the involved Knowledge Points and their shares are obtained. Consequently, an individual's learning history about a Knowledge Point can be generated. Figure 1 shows a typical learning history of a Knowledge Point. It records one's each learning experiences about a Knowledge Point. ``Learning cessation time" is used to calculate the interval between the learning time and current time, which can then be used to estimate how much information has been lost due to memory decay. ``Duration" is the length of a learning session. ``Proportion" is the Knowledge Point's share during a learning session.

\begin{figure}
	\centering
	\includegraphics[width=0.99\columnwidth]{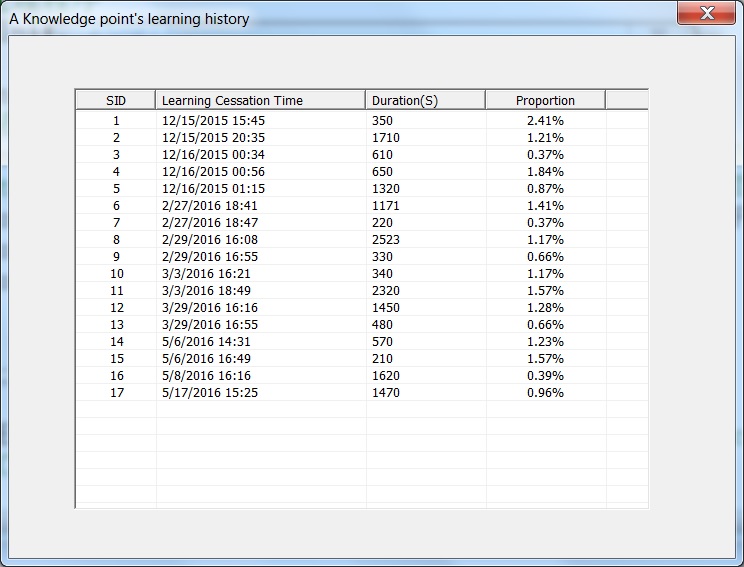}
	\caption{A typical learning history of a Knowledge Point}
\end{figure}
With the learning history, Equation 1 is utilized to calculate one's familiarity degree (called Familiarity Measure) about Knowledge Point  $ k_{i} $ at a particular time $ t $. The input is a sequence of $ m $ learning sessions (like Figure 1). $ d_{j} $ is session $ j $'s duration; $ \xi_{ij} $ is knowledge point $ k_{i} $'s share in session $ j $; $ b_{j} $ is the proportion of memory retention of learning session $ j $ at time $ t $, it is calculated with Ebbinghaus' Forgetting Curve equation\cite{ebbinghaus1913memory}.

	\begin{equation}
	F_{k_{i}} = \sum_{j=1}^md_{j}*\xi_{ij}*b_{j}
	\end{equation}
	
\subsection{Estimation of one's understanding of knowledge}
In Knowledge Model, Familiarity Measure is used as an estimation of one's knowing of a Knowledge Point, the knowing can be understanding it or just remembering it by rote. It is possible that a person is familiar with a Knowledge Point but does not understand it, because he does not know the background knowledge that is essential to understand it. This paper proposes a method to estimate whether an individual understands a Knowledge Point and how much he understands it. With the estimation, new applications can be facilitated, such as the one mentioned in Section 3.
\section{2. Understanding Tree}
If a person is familiar with a Knowledge Point and all the background knowledge that is essential to understand it, it is hypothesized that the person has understood the Knowledge Point; because Familiarity Measure depicts the cumulated effects of one's learning experiences about a topic, high Familiarity Measures imply intensive learning activities on the suite of Knowledge Points.

The background Knowledge Points that are essential to understand a Knowledge Point can be extracted by analyzing the definition of it. Table 1 lists eight reduced documents, each of them is a definition of a Knowledge Point in Probability Theory or Stochastic Process, the texts are quoted from Wikipedia and other websites. The third column of Table 1 lists the involved Knowledge Points in the documents, which are deemed as the background knowledge to understand the host Knowledge Point.

\begin{table*}
	\centering
	\begin{tabular}{|C{1cm}|L{12cm}|C{2.5cm}|}
		\hline
		Doc & \qquad \qquad \qquad \qquad \qquad \qquad \qquad \qquad Content & Knowledge points  \\ \hline
		D1 & A Strictly Stationary Process (SSP) is a Stochastic Process (SP) whose Joint Probability Distribution (JPD) does not change when shifted in time. &  SSP, JPD,\qquad\qquad Time, SP \\ \hline
		\qquad \qquad \qquad \qquad\qquad \qquad D2 &  A Stochastic Process (SP) is a Probability Model (PM) used to describe phenomena that evolve over time or space.  In probability theory, a stochastic process is a Time Sequence (TS) representing the evolution of some system represented by a variable whose change is subject to a Random Variation (RaV).  &  SP, PM, TS, Time, Space, System,\qquad \qquad Variable, RaV \\ \hline
		\qquad \qquad \qquad\qquad \qquad \qquad D3 &  In the study of probability, given at least two Random Variables (RV) X, Y, ... that are defined on a Probability Space (PS), the Joint Probability Distribution (JPD) for X, Y, ... is a Probability Distribution (PD) that gives the probability that each of X, Y, ... falls in any particular range or discrete set of values specified for that variable.  &  JPD, RV,\qquad\qquad PS, PD,\qquad \qquad Variable, Probability \\ \hline
		\qquad \qquad D4 &  A Probability model (PM) is a mathematical representation of a random phenomenon. It is defined by its Sample Space (SS), events within the SS, and probabilities associated with each event.  &  PM, SS,\qquad\qquad Event, Probability \\ \hline
		
		D5 &  In probability and statistics, a Random variable (RV) is a variable quantity whose possible values depend, in some clearly-defined way, on a set of random events.  &  RV, Variable, Event \\ \hline
		\qquad \qquad D6 &  A Probability Space (PS) is a Mathematical Construct (MC) that models a real-world process consisting of states that occur randomly. It consists of three parts: a Sample Space (SS), a set of events, and the assignment of probabilities to the events.  &  PS, MC, SS, Probability, Event \\ \hline
		D7 & A Probability Distribution (PD) is a table or an equation that links each outcome of a statistical experiment with its probability of occurrence. &  PD,\qquad\qquad Probability \\ \hline
		D8 & The Sample Space (SS) is the set of all possible outcomes of the samples. &  SS, Sample \\  	 
		\hline 
	\end{tabular}
	\caption{A list of documents and their involved Knowledge Points}
\end{table*}

An Understanding Tree is a treelike data structure which compiles the background Knowledge Points that are essential to understand the root Knowledge Point. Figure 2 illustrates four Understanding Trees based on the definitions of Table 1. The nodes of the tree can be further interpreted by other Knowledge Points until they are Basic Knowledge Points (BKP). A BKP is a Knowledge Point that is simple enough so that it is not interpreted by other Knowledge Points. Figure 3 shows a fully extended Understanding Tree based on the definitions of Table 1. Figure 4 shows an exemplary Understanding Tree with all the redundant nodes eliminated. Each node is tagged with a Familiarity Measure calculated with Knowledge Model. The leaf nodes of Figure 3 and Figure 4 are BKPs.

\begin{figure}
	\centering
	\includegraphics[width=0.99\columnwidth]{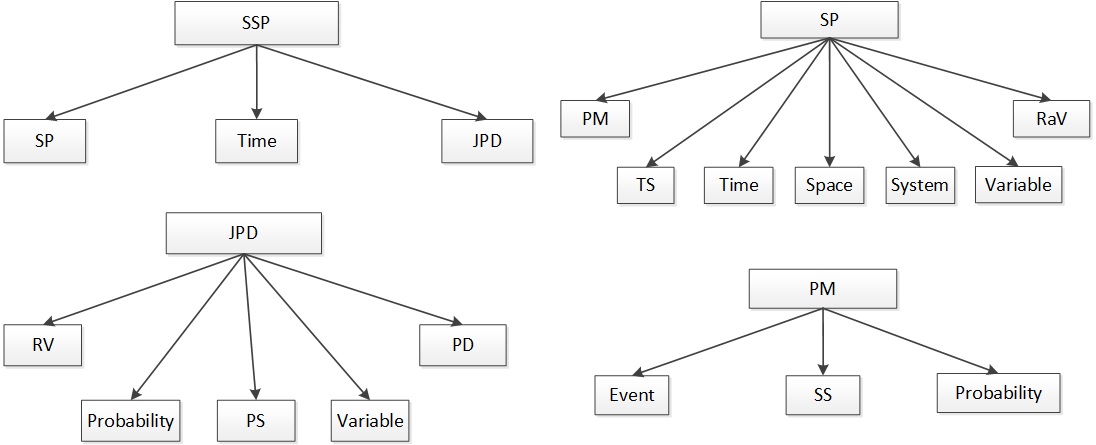}
	\caption{Some examples of Understanding Trees}
\end{figure}

\begin{figure*}
	\centering
	\includegraphics[width=1.8\columnwidth]{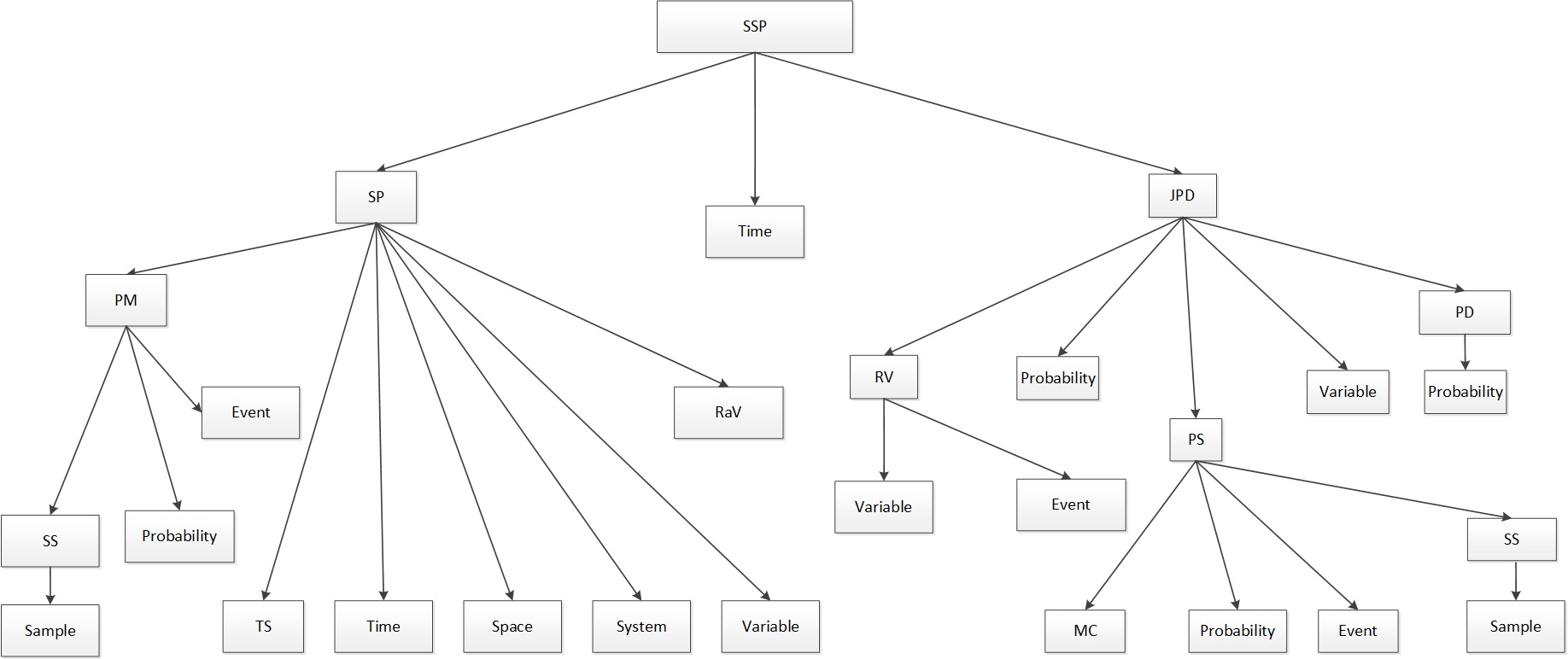}
	\caption{A fully extended Understanding Tree}
\end{figure*}

\begin{figure*}
	\centering
	\includegraphics[width=1.8\columnwidth]{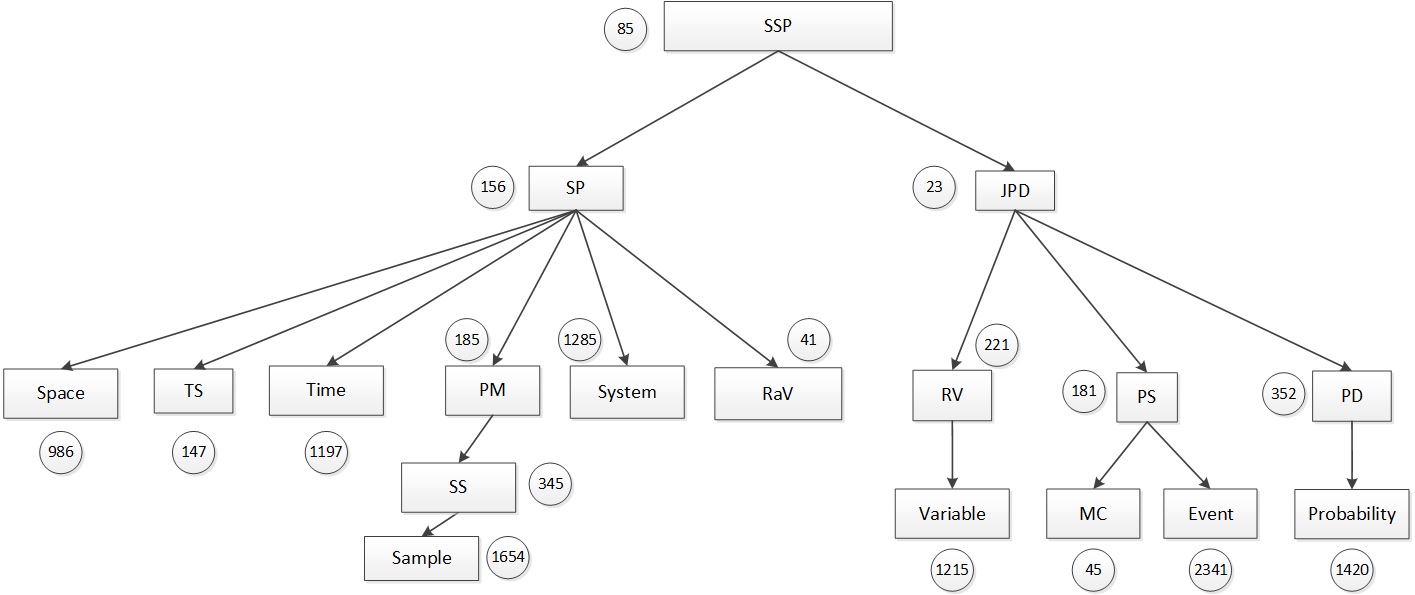}
	\caption{A standard Understanding Tree tagged with Familiarity Measures}
\end{figure*}
The height and number of nodes of an Understanding Tree characterize the complexity degree of the tree.
\subsection{Calculation of the quantity of understanding}
If all the Familiarity Measures of an Understanding Tree exceed a threshold (such as 100), it is assumed that the person has understood the root Knowledge Point. Due to the differences of people's intelligence and talent, different people may have different thresholds. If a Familiarity Measure is less than the threshold, a percentage is calculated by dividing it by the threshold, indicating the person's percent of familiarity of the node; if the Familiarity Measure is greater than the threshold, the percentage is set to 1. The person's understanding of the root Knowledge Point is calculated with Equation 2, $ PU $ is the percentage of understanding of the root, $ PF_{r} $ is the percentage of familiarity of the root, $ AP_{d} $ is the average percentage of familiarities of its descendants. Figure 5 is an exemplary Understanding Tree tagged with percentages, the $ PF_{r} $ of it equals 85\%, and the $ AP_{d} $ equals 89\%, so the $ PU $ equals 76\%, indicating the person's understanding of the root Knowledge Point is 76\%. If the $ PU $ is less than 1, the Knowledge Point is classified as ``Not Understood".

	\begin{equation}
	PU =   PF_{r} * AP_{d}
	\end{equation}
	
If all the percentages equal 1, the subject is assumed having understood the root Knowledge Point, then the average Familiarity Measure of the Understanding Tree features the magnitude of understanding. Since people are assumed to have well understood the BKPs, it may be preferable to exclude them or normalize their effects when computing the average Familiarity Measure. 
\begin{figure*}
	\centering
	\includegraphics[width=1.8\columnwidth]{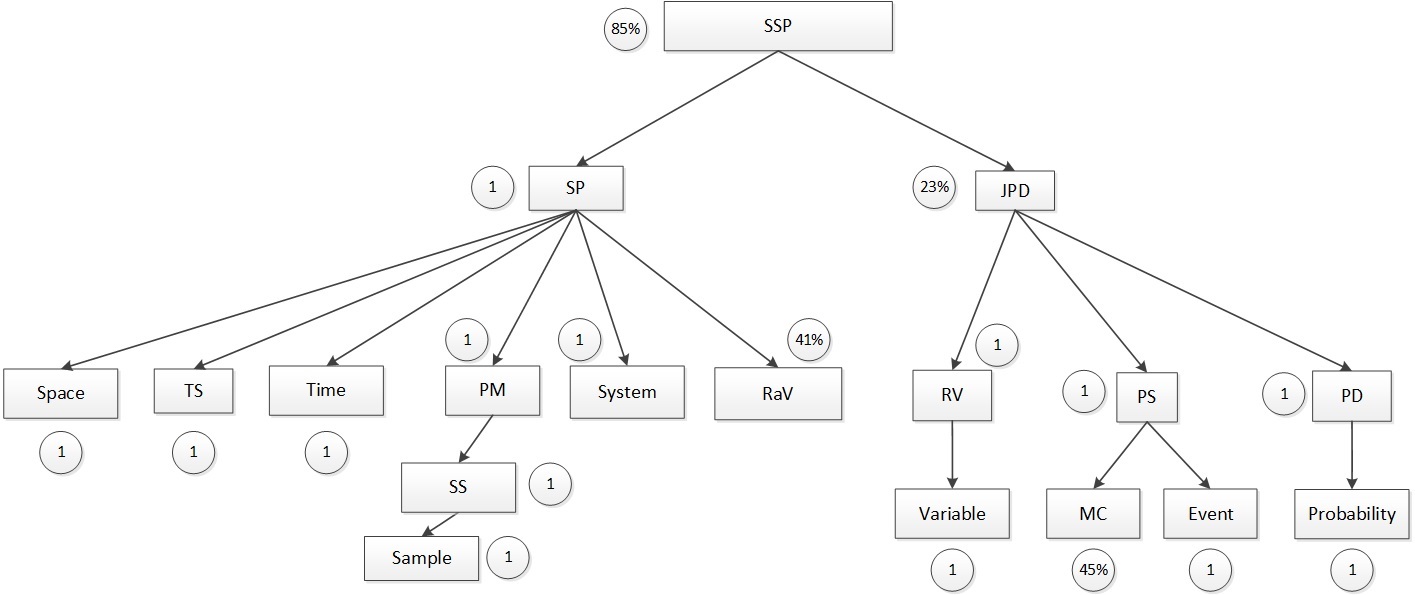}
	\caption{Familiarity Measures transformed into percentages}
\end{figure*}
\subsection{Construction of the Understanding Tree}
An Understanding Tree can be constructed manually by a group of experts, or generated automatically by machines. Algorithm 1 shows the steps to generate an Understanding Tree automatically. Table 2 illustrates three definitions of the Central Limit Theorem (CLT) extracted from several authoritative websites\footnote{\url{http://www.math.uah.edu/stat/sample/CLT.html}, \url{http://sphweb.bumc.bu.edu/otlt/MPH-Modules/BS/BS704_Probability/BS704_Probability12.html}, \url{http://stattrek.com/statistics/dictionary.aspx}}, the third column lists the involved Knowledge Points in each definition. According to the rule that a Knowledge Point must be present in more than half of the definitions, the child nodes of CLT are selected as \emph{sample}, \emph{distribution}, \emph{mean}, \emph{independent}, and \emph{normal} (Knowledge Points that are not BKPs can be further extended). In addition, the generated Understanding Tree can be inspected by human experts. The Understanding Tree is a static data structure, once constructed, they are unlikely to be altered. This characteristic makes the storage and retrieval of Understanding Trees convenient.

\begin{algorithm}[htb] 
	\caption{ An algorithm to construct Understanding Tree}  
	\begin{algorithmic}[1] 
		\REQUIRE ~~\\ 
		
		The set of all knowledge points, $ \Omega $;\\
		The set of all BKPs, $ B $;\\
		The root knowledge point, $ R_k $;
		
		\ENSURE ~~\\ 
		$ R_k $'s Understanding Tree;
		\STATE Search definitions of $ R_k $ from a library of authoritative documents;
		\STATE Discover involved knowledge points for each definition;
		\STATE Select knowledge points according to some rules, e.g., more than half of the definitions have referenced the knowledge point; 
		\STATE Recursively extend non-BKP knowledge points that are selected;
		\RETURN All the selected Knowledge Points;		
	\end{algorithmic}
\end{algorithm} 

\begin{table*}
	\centering
	\begin{tabular}{|C{1cm}|L{12cm}|C{3.5cm}|}
		\hline
		  & \qquad \qquad \qquad \qquad \qquad \qquad \qquad \qquad Content & Knowledge Points  \\ \hline
		\qquad \qquad \qquad \qquad \qquad \qquad 1 & The Central Limit Theorem (CLT) states that the sampling distribution of the mean of any independent, random variable will be normal or nearly normal, if the sample size is large enough. &  CLT,  sample, distribution, mean, independent, random variable, normal, size \\ \hline
	    \qquad \qquad \qquad \qquad \qquad \qquad 2 & The Central Limit Theorem (CLT) states that the distribution of the sum (or average) of a large number of independent, identically distributed variables will be approximately normal, regardless of the underlying distribution.  &  CLT,  distribution, sum,  average, independent, variable, normal \\ \hline
		\qquad \qquad \qquad \qquad \qquad \qquad 3 &  The Central Limit Theorem (CLT) states that if you have a population with mean $ \mu $ and standard deviation $ \sigma $ and take sufficiently large random samples from the population with replacement, then the distribution of the sample means will be approximately normally distributed.  &  CLT,  population, standard deviation, random, replacement, distribution, sample, mean, normal \\  
		\hline 
	\end{tabular}
	\caption{Three definitions of the Central Limit Theorem (CLT)}
\end{table*}

\subsection{The Reverse Understanding Tree}
The reverse tree of Figure 3 is used to discover critical Knowledge Points that are heavily relied on by other Knowledge Points. Figure 6 shows three reverse Understanding Tree based on the tree of Figure 3, they reflect dependency relationships of Knowledge Points.

\begin{figure}
	\centering
	\includegraphics[width=0.99\columnwidth]{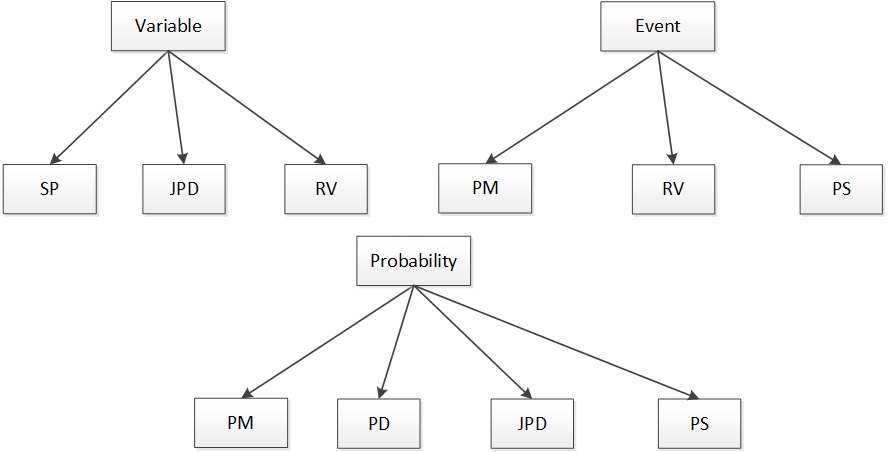}
	\caption{Some examples of Reverse Understanding Trees}
\end{figure}
\section{3. Computer-aided meaningful learning}
Knowing the quantities of one's understanding of all the Knowledge Points facilitates many new applications, besides the use cases mentioned in \cite{Liu2016}, it can augment Meaningful Learning. Meaningful learning is the concept that learned knowledge is fully understood to the extent that it relates to other knowledge, implies there is a comprehensive knowledge of the context of the facts learned\cite{Dadic2008}.  Computer-aided incremental meaningful learning (CAIML) states a strategy to let the computer estimates a person's current knowledge states and knowledge compositions, then plans the optimum learning material for the person to accomplish a Meaningful Learning. The optimum material introduces some new knowledge blended with old knowledge the subject has known, meanwhile, interpreting the new knowledge.

For example, a college student, who has comprehended the basic knowledge of advanced mathematics and Computer Science, wants to be an expert of Artificial Intelligence (AI) by teaching himself. A professor recommended him 1000 documents related to AI, and asserted if he can fully understand the documents, he will be an expert of AI. The question is: what is the best sequence for the student to learn the documents? Some documents are easy to understand, and should be read in the beginning; some documents are intricate, learning them in the first place is painful and frustrating, and should be put at the end of the learning process. If a computer knows a person's knowledge states at any time, it is not difficult for it to make the learning plan, and recommend a document for current learning.
\subsection{An algorithm to facilitate CAIML}
Algorithm 2 is devised to facilitate CAIML, it recommends the optimum document for the subject to learn, by analyzing his current knowledge states. It searches for the document which has the least Knowledge Points that are not understood by the subject, which implies it is the easiest document to understand at present. In the document, the ``Understood" Knowledge Points server as interpretations to the ``Not Understood" ones when learning the document, therefore, the algorithm facilitates a Meaningful Learning.

\begin{algorithm}[htb] 
	\caption{ An algorithm to recommend the optimum document(s) for CAIML}  
	\begin{algorithmic}[1] 
		\REQUIRE ~~\\ 
		
		The set of documents to be learned;\\
		The set of all Knowledge Points, tagged with the subject's Familiarity Measures about them;\\
		The set of all non-BKP Knowledge Points' Understanding Trees;
		
		\ENSURE ~~\\ 
		The optimum document(s) for current learning to practice a CAIML;
		\STATE Extract involved Knowledge Points in each document, classify them as ``Understood" and ``Not Understood", according to the rule mentioned in Section 2;
		\STATE Count the number of ``Not Understood" Knowledge Points for each document, exclude documents that all involved Knowledge Points are ``Understood";
		\RETURN The document(s) with the least ``Not Understood" Knowledge Points;		
	\end{algorithmic}
\end{algorithm}

An alternative method to recommend the optimum document is to estimate a person's current understanding of each document, then return the one that is most approaching 100\% (but does not equal 100\%). A person's understanding of a document is calculated with Equation 3. Supposing the document involves $ n $ Knowledge Points, $ PUD $ is the percentage of understanding of the document; $ \phi_{k} $ is Knowledge Point $ k $'s share of the document, its calculation refers to Formula 2 of \cite{Liu2016}; $ PU_{k} $  is Knowledge Point $ k $'s percentage of understanding. If a person has understood all the Knowledge Points the document contains, its PUD is 100\%. It is possible that a person learns a ``100\% understood" document to strengthen his knowledge.

\begin{equation}
  PUD  = \frac{\sum_{k=1}^n\phi_{k}*PU_{k}}{\sum_{k=1}^n\phi_{k}}
\end{equation}

People's memory fades away over time, the Familiarity Measures decrease accordingly. It is also possible that the computer recommends a document that had been tagged ``100\% understood", because the PUD has abated.
\subsection{An example of CAIML}
Here is an example to illustrate the logics of CAIML. A person wants to fully understand the suite of documents listed in Table 1 by learning them. Figure 4 shows all the Knowledge Points referenced by the documents, the third column of Table 1 lists ones referenced by each of them. The subject is assumed to have understood the BKPs before the beginning of the learning, which are the leaf nodes of Figure 4; if not, he can learn them first. Table 3 shows the number of ``Not Understood" Knowledge Points before the beginning of the learning and after each learning, for each document, e.g., there are 8 ``Not Understood" Knowledge Points before the starting of the learning for D1, they are SSP, SP, JPD, PM, SS, RV, PS, and PD. According to Algorithm 2, the subject should first learn either of D5, D7 or D8. After learning of D5, the subject is assumed to have understood RV, therefore, the number of ``Not Understood" Knowledge Points for D1 becomes 7. The first column of Table 3 suggests one of the optimum learning sequences of the documents, which is D5, D8, D4, D2, D7, D6, D3, and D1.

\begin{table}
	\centering
	\begin{tabular}{|C{1.5cm}| c| c| c| c| c| c| c| c|}
		\hline
		Learning sequence & D1 & D2 & D3 & D4 & D5 & D6 & D7 & D8\\ \hline
		Before starting & 8 & 3 & 5 & 2 & 1 & 2 & 1 & 1\\ \hline
		D5 & 7 & 3 & 4 & 2 & 0 & 2 & 1 & 1\\ \hline
		D8 & 6 & 2 & 3 & 1 & 0 & 1 & 1 & 0\\ \hline
		D4 & 5 & 1 & 3 & 0 & 0 & 1 & 1 & 0\\ \hline
		D2 & 4 & 0 & 3 & 0 & 0 & 1 & 1 & 0\\ \hline
		D7 & 3 & 0 & 2 & 0 & 0 & 1 & 0 & 0\\ \hline
		D6 & 2 & 0 & 1 & 0 & 0 & 0 & 0 & 0\\ \hline
		D3 & 1 & 0 & 0 & 0 & 0 & 0 & 0 & 0\\ \hline
		D1 & 0 & 0 & 0 & 0 & 0 & 0 & 0 & 0\\   	 
		\hline 
	\end{tabular}
	\caption{An example of CAIML}
\end{table}

\section{4. Discussion}
\subsection{The difficulty levels of Knowledge Points}
One of Knowledge Model's limitations is that it needs a mechanism to discriminate the difficulty level of each Knowledge Point, so that the calculated Familiarity Measures can be normalized among Knowledge Points, then the Familiarity Measures can be compared between Knowledge Points. 
With Understanding Tree, it is not necessary to discriminate the difficulty levels of Knowledge Points, because a Knowledge Point's magnitude of complexity can be reflected by the size of its Understanding Tree; a complicated Knowledge Point corresponds to a larger tree, a simple one corresponds to a smaller tree.
\subsection{Other methods of generating learning histories and calculating Familiarity Measures}
There are other methods of generating learning histories and calculating Familiarity Measures, such as counting the occurrence frequency of each Knowledge Point experienced by a person and using the frequency as a Familiarity Measure.
Maybe one day wearable computers can conveniently know what a person is thinking by analyzing his brainwaves. Knowing what actually happened in one's brain, instead of inferring, can generate a more accurate learning history. Further research is necessary to evaluate the performance of different methods.
\subsection{Knowledge Points that are different but have little distinction}
There is a problem of how to deal with Knowledge Points that are different but have little distinction, such as ``random variation" and ``random variable". One solution is to homogenize them to the same Knowledge Point; another solution is to compensate one Knowledge Point's Familiarity Measure with others' Familiarity Measures, because learning others helps to understand it. The compensation can be calculated with Equation 4. $ F_{k_{i}} $ is Knowledge Point $ i $'s Familiarity Measure, each of its sibling contributes $ 1/k $ of its Familiarity Measure to $ i $ ($ k $ is a coefficient to be determined). 

	\begin{equation}
	F_{k_{i}\_new} = F_{k_{i}\_old} + \sum_{j=1}^m{\frac{1}{k}}F_{k_{j}}
	\end{equation}

\subsection{Other potential applications of Knowledge Model and Understanding Tree}
Knowing one's possession or understanding of knowledge, a lot of new applications become possible. Such as discovering a person's deficiencies in a field, so he can remedy the deficiencies; gathering people sharing the same interest of topics; analyzing the knowledge states of geniuses when they accomplished their masterpieces, with the techniques of Big Data Analysis, maybe some patterns can be found. Similarly, other categories of people's knowledge states can be analyzed, such as criminals.
\subsection{Trade-offs of evaluating one's possession of knowledge}
Quantitatively assessing one's knowledge seems to be a good thing, but there are risks that it introduces some harmful effects. For example, if the Familiarity Measures calculated are inaccurate, it may lead to wrong decisions. In addition, it cannot detect a person's talent and potential in a field. On the other hand, traditional exams or interviews have their limitations. For example, it needs other people's cooperation to accomplish the evaluation; it can only assess one's knowledge in a particular field at a time, and the evaluation is not comprehensive, because it only assesses questions being asked. Knowledge Model and Understanding Tree can assess one's knowledge solitarily, automatically, and comprehensively. Therefore, the methods of evaluating one's possession of knowledge should be used cooperatively, complementing one another.
\section{5. Related work}
Many research fields focus on the collection of personal information, such as lifelogging, expertise finding, and personal informatics.
Personal informatics is a class of tools that help people collect personally relevant information for the purpose of self-reflection and gaining self-knowledge\cite{li2010stage,wolf2009know,yau2009self}. Various tools have been developed to help people collect and analyze different kinds of personal information, such as location\cite{lindqvist2011m}, finances\cite{kaye2014money}, food\cite{cordeiro2015barriers}, weight\cite{kay2013there,maitland2011designing}, and physical activity\cite{fritz2014persuasive}. Knowledge Model and Understanding Tree facilitate a new type of personal informatics tool that helps people discover their expertise and deficiencies in a more accurate way, by quantitatively assessing an individual's possession of knowledge.

Expertise is one's expert skill or knowledge in a particular field. Expertise finding is the use of tools for finding and assessing individual expertise\cite{mcdonald1998just,mattox1999enterprise,vivacqua1999agents}. As an important link of knowledge sharing, expertise finding has been heavily studied in many research communities\cite{ackerman2013sharing, white2009characterizing,maybury2002awareness,tang2008arnetminer,pipek2003pruning,guy2013mining}. Many sources of data have been exploited to assess an individual's expertise, such as one's publications, documents, emails, web search behavior, other people's recommendations, social media etc. Knowledge Model and Understanding Tree provide a new source of data to analyze one's expertise -- one's learning history about a subject, which is more comprehensive and straightforward than other data sources, because one's expertise is mainly obtained through learning (Including ``Informal Learning", which occurs through the experience of day-to-day situations, such as a casual conversation, play, exploring, etc.)

\section{6. Conclusion}
This paper and \cite{Liu2016} propose a framework of evaluating a person's possession of knowledge by keeping track of his/her learning activities. \cite{Liu2016} proposes a computation framework to calculate Familiarity Measures, this article devises a data structure called Understanding Tree to estimate one's understanding of a Knowledge Point, based on the Familiarity Measures. With the prevailing of wearable computers like Google Glass and Apple Watch, and maturing of technologies like Speech Recognition and Optical Character Recognition (OCR), it is not difficult to analysis people's daily learning activities like talking, listening, and reading. In addition, the capabilities of computers are growing rapidly, it is practicable to record and analyze a person's learning activities continually. Even if we cannot record all of one's learning activities, a sample of them also help to estimate one's possession of knowledge. For most applications proposed in this paper and \cite{Liu2016}, some deviation of estimation is tolerable and can be adjusted by the users. Therefore, the framework is technically feasible and beneficial.

\balance{}

\bibliographystyle{SIGCHI-Reference-Format}
\bibliography{kmodel}

\end{document}